\documentclass[
notitlepage
,floatfix,
aps,
pra,
reprint,
superscriptaddress,
]{revtex4-1}
\usepackage[utf8]{inputenc}

\usepackage[caption=false]{subfig}
\usepackage{graphicx} 
\usepackage{epstopdf}
\usepackage{array}
\usepackage{verbatim}
\usepackage{amsmath,amsfonts,amssymb,amscd,mathtools}
\usepackage{amsthm}
\usepackage{tabularx}
\usepackage{stmaryrd}
\usepackage{enumerate}
\usepackage[ruled]{algorithm2e}
\usepackage{wasysym}
\usepackage{braket}
\usepackage{mathrsfs}
\usepackage{microtype}
\usepackage{hyperref}
\usepackage[dvipsnames]{xcolor}
\hypersetup{
    bookmarksnumbered=true, 
    unicode=false, 
    pdfstartview={FitH}, 
    pdftitle={}, 
    pdfauthor={}, 
    pdfsubject={}, 
    pdfcreator={}, 
    pdfproducer={}, 
    pdfkeywords={}, 
    pdfnewwindow=true, 
    colorlinks=true, 
    linkcolor=NavyBlue, 
    citecolor=NavyBlue, 
    filecolor=NavyBlue, 
    urlcolor=NavyBlue 
}

\theoremstyle{plain}
\newtheorem{thm}{Theorem}
\newtheorem{cor}[thm]{Corollary}
\newtheorem{lem}[thm]{Lemma}
\newtheorem{pro}[thm]{Proposition}

\theoremstyle{definition}

\newcommand{\eq}[1]{(\hyperref[eq:#1]{\ref*{eq:#1}})}

\renewcommand{\sec}[1]{\hyperref[sec:#1]{Section~\ref*{sec:#1}}}
\newcommand{\thrm}[1]{\hyperref[thrm:#1]{Theorem~\ref*{thrm:#1}}}
\newcommand{\lemm}[1]{\hyperref[lemm:#1]{Lemma~\ref*{lemm:#1}}}
\newcommand{\prop}[1]{\hyperref[prop:#1]{Proposition~\ref*{prop:#1}}}
\newcommand{\corr}[1]{\hyperref[corr:#1]{Corollary~\ref*{corr:#1}}}
\newcommand{\fig}[1]{\hyperref[fig:#1]{~\ref*{fig:#1}}}
\newcommand{\deff}[1]{\hyperref[deff:#1]{~\ref*{deff:#1}}}

\newcommand{\mU}{\mathcal{U}}

\newcommand{\mD}{\mathcal{D}}
\newcommand{\mI}{\mathcal{I}}
\newcommand{\mF}{\mathcal{F}}
\newcommand{\mL}{\mathcal{L}}

\newcommand{\mG}{\mathcal{G}}
\newcommand{\mH}{\mathcal{H}}
\newcommand{\mM}{\mathcal{M}}

\newcommand{\mbI}{\mathbb{I}}

\newcommand{\mfD}{\mathfrak{D}}

\newcommand{\cleq}{\preceq}
\newcommand{\cgeq}{\succeq}

\DeclareMathAlphabet{\matheu}{U}{eus}{m}{n}

\DeclareMathOperator{\Tr}{Tr}
\DeclareMathOperator{\id}{id}

\newcommand{\ketbra}[2]{|{#1}\rangle\!\langle{#2}|}

\newcommand{\ba}{\begin{eqnarray}}
\newcommand{\ea}{\end{eqnarray}}
\newcommand{\bann}{\begin{eqnarray*}}
\newcommand{\eann}{\end{eqnarray*}}
\newcommand{\bal}{\begin{equation}\begin{aligned}}
\newcommand{\eal}{\end{aligned}\end{equation}}
\newcommand{\dm}[1]{\ketbra{#1}{#1}}

\newcolumntype{L}[1]{>{\raggedright}p{#1}}
\newcolumntype{C}[1]{>{\centering}p{#1}}
\newcolumntype{R}[1]{>{\raggedleft}p{#1}}
\newcolumntype{D}{>{\centering\arraybackslash}X}

\newcommand{\sbar}{\;\rule{0pt}{9.5pt}\right|\;}
\newcommand{\lset}{\left\{\left.}
\newcommand{\rset}{\right\}}

\begin{document}
\title{Universal limitations on implementing resourceful unitary evolutions}

\author{Ryuji Takagi}
\email{rtakagi@mit.edu}
\affiliation{Center for Theoretical Physics and Department of Physics, Massachusetts Institute of Technology, Cambridge, Massachusetts 02139, USA}
\author{Hiroyasu Tajima}
\email{hiroyasu.tajima@yukawa.kyoto-u.ac.jp}
\affiliation{Yukawa Institute for Theoretical Physics, Kyoto University,
Oibuncho Kitashirakawa Sakyo-ku, Kyoto, 606-8502, Japan}

\begin{abstract}
We derive a trade-off relation between the accuracy of implementing a desired unitary evolution using a restricted set of free unitaries and the size of the assisting system, in terms of the resource generating/losing capacity of the target unitary. In particular, this relation implies that, for any theory equipped with a resource measure satisfying lenient conditions, any resource changing unitary cannot be perfectly implemented by a free unitary applied to a system and an environment if the environment has finite dimensions. Our results are applicable to a wide class of resources including energy, asymmetry, coherence, entanglement, and magic, imposing ultimate limitations inherent in such important physical settings, as well as providing insights into operational restrictions in general resource theories. 
\end{abstract}

\maketitle

\section{Introduction}

One of the ultimate goals in quantum information science is to understand the operational enhancement made possible by quantum phenomena as well as limitations on the enhancement imposed by laws of quantum mechanics. 
This is not only an important theoretical question but also of practical relevance, as recent years have witnessed the burgeoning development in manipulation of systems on small scales, in which quantum effects play central roles.

Any quantum information processing involves time evolution of quantum states, and the most fundamental building block for the quantum dynamics is unitary evolution.
Even though general quantum dynamics is described by completely positive trace preserving (CPTP) maps, also called quantum channels, any channel acting on a system can be simulated by an appropriate unitary operation applied over the system and an environment \cite{Stinespring1955}, and thus any quantum evolution can be realized if one has access to an arbitrary unitary. 
However, due to technological limitations as well as restrictions imposed by laws of physics, physical systems usually do not allow one to apply an arbitrary unitary. This makes it essential to consider to what extent a desired unitary dynamics can be realized only using a limited set of accessible unitaries.
This question has been specifically addressed for the systems with additive conserved quantities, in which only unitaries that respect the conservation laws can be applied \cite{Ozawa2002conservative,Ozawa2003uncertainty,Karasawa2007conservation,Karasawa2009gate,Aberg2014catalytic,Tajima2018uncertainty, Tajima2019coherence}. 
In particular, Ref. \cite{Tajima2018uncertainty} has derived a lower bound for the necessary amount of quantum fluctuation that the ancillary state must possess to implement a desired unitary in terms of its implementation accuracy and the amount of energy that the target unitary can create, and they further derived lower and upper bounds that always match asymptotically in the region where the implementation error is small \cite{Tajima2019coherence}.  
The presented bounds lead to a fundamental no-go theorem that prohibits the perfect implementation of any unitary that can create energy using an energy conserving unitary and finite-sized ancillary state.

However, there are various settings where other types of quantities can play the main role, and one can ask whether this type of trade-off relation is a general property shared by generic physical situations. 
This line of thought naturally leads to the idea of resource theories, which are general frameworks that deal with quantification and manipulation of precious quantities considered ``resource" under a given setting \cite{Chitambar2019resource}.
The resource theoretic framework allows for systematic investigation on specific physical settings \cite{plenio2007introduction,HOrodecki_review2009,aberg2006quantifying,Baumgratz2014,Streltsov2017,Theurer2019quantifying,Gour2008,Marvian2016,Brandao2013,Brandao_secondlaws2015,gallego_2015,Rivas2010non-Markov,wakakuwa2017operational,Veitch_2014stab,howard_2017,Genoni2008,Takagi2018,albarelli2018resource} and has turned out to be especially useful for providing a unifying operational view to general class of quantities \cite{horodecki_2012,brandao_2015,delrio_2015,coecke_2016,Liu2017,gour_2017,anshu_2017,regula_2018,Lami2018gaussian,takagi2018operational,li_2018,Takagi2019general,uola2018quantifying,Liu2019oneshot,Vijayan2019oneshot}. 
In this context, it can be seen that the previous works \cite{Tajima2018uncertainty,Tajima2019coherence} dealt with a specific theory (i.e. theory of asymmetry with U(1) group \cite{Gour2008,Marvian2016}), and it has remained elusive whether one can extend the relevant consideration to more general resources.

Here, we address the above question for the setting where a set of ``free" (i.e. accessible) unitaries is given, and one aims to implement ``resourceful" (i.e. non-free) unitaries with a free unitary and an aiding state defined in the ancillary system. 
Our main results are trade-off relations between the implementation accuracy, the amount of resources that the target unitary can change, and the size of the ancillary system, which are applicable to a wide class of physical settings that satisfy several lenient conditions. 
These relations immediately lead to no-go theorems that prohibit us from implementing any resourceful unitary with perfect accuracy only using free unitaries and aiding states defined in a system with finite size, which qualitatively reproduces the results in \cite{Tajima2018uncertainty,Tajima2019coherence} as a special case. 
We also apply our results to several important settings and discuss significance of the results. 

This paper is organized as follows. In Section \ref{sec:free unitaries}, our setup and useful quantities as well as conditions that play major roles in later discussions are introduced. In Section \ref{sec:implementation}, our first main result on the trade-off relation between accuracy, the amount of resources the target unitary can change, and the size of the ancillary system is presented. 
In Section \ref{sec:relaxation}, we show our second main result that relaxes one of the conditions in the trade-off relation, which significantly increases its applicability. 
In Section \ref{sec:application}, we apply our results to various resources such as energy, asymmetry, coherence, entanglement, and magic. 
In Section \ref{sec:full generality}, we discuss possibilities of extending the no-go result to even more general settings. 
We finally conclude our discussion in Section \ref{sec:conclusions}.

\section{Free unitaries and resource measures}\label{sec:free unitaries}
Let $\mH_d$ denote the Hilbert space with dimension $d$ and $\mD(\mH_d)$ be the set of density operators acting on $\mH_d$. 
Also, let $\mU_\mF(d)\subseteq U(d)$ be some set of unitaries acting on $\mH_d$ and define $\mU_\mF:=\bigcup_d \mU_\mF(d)$, which we call a set of \textit{free unitaries}.    
The set of free unitaries is usually determined by the system of interest, and it can be most naturally understood as free operations in the context of resource theories. 
A resource theory is specified by its set of free states and free operations, which are considered given for free under the interested physical setting, and an important requirement for free operations is that they are not capable of creating any resources out of free states. For instance, for the setting where two parties are physically separated apart, a reasonable theory comes with the set of separable states as free states and the set of local operations and classical communication (LOCC) as free operations.
Motivated by the resource theoretic considerations, we also define \textit{resource measures} as the maps from states to non-negative real numbers. 
If one assumes some underlying resource theory of quantum states, one natural choice is to take resource monotones (which evaluate zero for free states and do not increase under application of free operations) defined in the theory as resource measures.

Once some resource theory is provided, one can naturally consider $\mU_\mF$ as the set of unitaries that are also free operations (e.g. the set of local unitaries for the case of entanglement.) 
However, although considering the underlying resource theory is conceptually useful, for our purpose as long as the set of free unitaries is given, one does not necessarily need to assume an underlying structure of the resource theory.
Indeed, as we shall see later it is sometimes convenient to only consider the set of free unitaries, not explicitly taking into account the underlying set of free states. 
In the same vein, we do not impose the monotonicity property for resource measures in general. 
Instead, we consider the following properties for a resource measure $R$ determined by the given set of free unitaries, which play major roles in later discussions. 

\begin{description}
\item[Property 1]{(Invariance under free unitaries) $R(\rho)= R(V\rho V^\dagger),\ \forall V\in \mU_\mF$.}
\item[Property 2]{(Continuity) There exist non-negative increasing functions $f$, $g$ with $\lim_{x\rightarrow0}f(x)=0$, $g(x)<\infty,\ \forall x<\infty$, and a real function $h$ with $\lim_{x\rightarrow0}h(x)=0$ such that 
\begin{align}
|R(\rho)-R(\sigma)|\le f(D(\rho,\sigma))g(d)+h(D(\rho,\sigma))
\label{eq:continuity}
\end{align}
for $\rho,\ \sigma\in \mD(\mH_d)$ where $D(\rho,\sigma)$ is some distance measure between $\rho$ and $\sigma$. 
}
\item[Property 3]{(Additivity for product states) $R(\rho\otimes\sigma)=R(\rho)+R(\sigma)$.}
\end{description}

Property~1 refers to the fact that free unitaries do not change the resource contents attributed to quantum states, and it is especially a natural property when application of a free unitary can be reversed by another free unitary.
Property~2 states that if two states are close to each other, the amount of resources possessed by these states should be also close. 
Property~3 is the property that if a state is prepared independently of another state, the resource contents attributed to the two states are evaluated as the sum of the amount of resources possessed by each state.
As we see in Section~\ref{sec:application}, these properties are shared by a number of known resource measures, and we shall obtain ultimate bounds on implementation accuracy of desired unitary in terms of the resource measures satisfying these conditions.

We also define the \textit{resource generating power} and \textit{resource losing power} for unitary $U$ \cite{zanardi_2000,Diaz2018usingreusing,seddon2019quantifying,Wang2019magic,li_2018}:
\begin{align}
\mG_{U}&:=\max_{\rho} \left\{R(U\rho U^\dagger)-R(\rho)\right\},\label{eq:resource generating}\\
\mL_{U}&:=-\min_{\rho} \left\{R(U\rho U^\dagger)-R(\rho)\right\}.\label{eq:resource losing}
\end{align}
Note that $\mG_U,\mL_U\geq 0$ for any $U$ because there always exists a state $\rho$ that is invariant under $U$, for which one can for instance take $\rho=\dm{u}$ where $\ket{u}$ is an eigenstate of the unitary.

\section{Implementation of resourceful unitaries} \label{sec:implementation}

Once the concept of free unitaries is introduced, one can ask what can be done with them and what are ultimate limitations imposed on the tasks accomplished by the given free unitaries. 
One of the fundamental questions that is both practically and theoretically important is whether we can implement (or simulate) non-free unitaries, which we call \textit{resourceful unitaries}, only using free unitaries with the aid of the ancillary system.  

More specifically, our aim is to simulate the given unitary $U_S$ on the Hilbert space $\mH_S$ by a channel $\Lambda_S$ implemented by a free unitary $V_{SE}\in\mU_\mF$ acting on the Hilbert space $\mH_S\otimes \mH_E$ and some ancillary state $\rho_E\in \mD(\mH_E)$, i.e.
\ba
 \Lambda_S(\cdot):=\Tr_{E}[V_{SE}(\cdot\otimes\rho_E)V_{SE}^{\dagger}].
 \label{eq:implementation with free unitary}
\ea

The tuple $\mI:=(\mH_E,V_{SE},\rho_E)$ defines a specific implementation of the channel.
A standard way of evaluating the closeness of two quantum channels is to see how close the output states from these channels are when the channels are allowed to act on only part of the input space. 
In order to take into account the worst-case input, we define the error for the given implementation $\mI$, which is a type of gate fidelity, as 
\begin{align}
\delta_\mI^{U_S}:=\max_{\rho_S}\delta_\mI^{U_S}(\rho_S)
\end{align}
where
\ba
 \delta_\mI^{U_S}(\rho_S)&:=&L_{e}(\rho_S,\Lambda_{U^{\dagger}_S}\circ\Lambda_S),\\
 \Lambda_{U}(\cdot)&:=& U\cdot U^\dagger 
\label{eq:accuracy bures}
\ea
and
\begin{align}
  L_{e}(\rho_S,\Lambda )  &:=\sqrt{2(1-F_{e}(\rho_S,\Lambda))},\label{uerror}\\
  F_{e}(\rho_S,\Lambda)  &:=\sqrt{\bra{\psi}_{SR}  [\Lambda\otimes \id_R](\psi_{SR})  \ket{\psi}_{SR}}.
\end{align}
where $\ket{\psi}_{SR}$ is a purification of $\rho_S$.
A related distance measure is the Bures distance for two quantum states:
\bal
 L(\rho,\sigma):=\sqrt{2\left(1-F(\rho,\sigma)\right)}
\eal
 where $F(\rho,\sigma):=\|\sqrt{\rho}\sqrt{\sigma}\|_1$ is the Uhlman fidelity.
The choice of this distance measure is primarily due to the mathematical convenience in later discussions, but because of the well-known relations with other distance measures, one can easily transform the results to the ones with respect to other measures as well --- indeed, we will reformulate the relation in terms of the distance measure based on trace norm and diamond norm, which come with clear operational meaning in terms of distinguishability. 

Then, we obtain the following trade-off relation between resourcefulness of desired unitary, implementation accuracy, and dimension of the ancillary system with respect to \textit{any} resource measure satisfying the three properties above. 

\begin{thm}\label{thm:sizebound}
Let $R$ be a resource measure satisfying Properties 1, 2, 3 and $f_L$, $g_L$, $h_L$, $\mG_{U_S}$, $\mL_{U_S}$ be the functions defined in \eqref{eq:continuity}, \eqref{eq:resource generating}, \eqref{eq:resource losing} with respect to $R$ and the Bures distance: $D(\rho,\sigma):=L(\rho,\sigma)$.    
Then, for any implementation $\mI$, it holds that 
\begin{align}
\mG_{U_S}+\mL_{U_S}&\le \alpha_L(\delta_\mI^{U_S},d_E)+\beta_L(\delta_\mI^{U_S}).
\end{align}
where $\alpha_L(x,y):=f_L(2\sqrt{2}x)g_L(y)+2f_L(2x)g_L(d_S\cdot y)$, $\beta_L(x):=h_L(2\sqrt{2}x)+2h_L(2x)$ with $d_E:=\dim\mH_E$, $d_S:=\dim\mH_S$.
\end{thm}
The proof of Theorem \ref{thm:sizebound} can be concisely stated by utilizing the ``no-correlation lemma" shown in \cite{Tajima2019coherence}, which quantitatively clarifies the fact that in order to implement a unitary on the target system approximately, the correlation between the target system and the external device must become weak. 
We defer a detailed proof to the Appendix. 
Note that $\alpha_L$ and $\beta_L$ are increasing functions that approach 0 as $x, y\rightarrow0$.
Thus, fixing the dimension of the system of interest, Theorem \ref{thm:sizebound} can be seen as a trade-off relation between the size of the device in the ancillary system and the implementation accuracy, and in particular the result indicates that in order to implement a resourceful unitary the dimension of the ancillary system must grow as the implementation becomes better, and at the limit of perfect implementation the size of the ancillary system must diverge. 
Notably, Theorem \ref{thm:sizebound} holds for \textit{any} resource measure that satisfies Properties 1, 2, 3, which ensures a wide applicability of the trade-off relation. 
This observation immediately leads to the following fundamental no-go theorem.

\begin{cor}\label{cor:no-go perfect}
 Given the set of free unitaries $\mU_\mF$ and a finite dimensional ancillary system $\mH_E$ with $\dim\mH_E<\infty$, it is impossible to perfectly implement any unitary that can generate (or lose) nonzero resources in terms of at least one resource measure satisfying Properties 1, 2, 3 by means of Eq. \eqref{eq:implementation with free unitary}.  
\end{cor}

Theorem~\ref{thm:sizebound} and Corollary~\ref{cor:no-go perfect} suggest an important implication --- one might think that if a target operation can only create a certain amount of resource, supplying a state defined in a finite-dimensional space with roughly the same amount of resource would be enough to accomplish the desired implementation. The above results state that it is not the case when it comes to the unitary implementation, and Theorem~\ref{thm:sizebound} in particular provides a quantitative estimation of the necessary dimension even when a non-zero error is allowed.  

It is also convenient to rewrite Theorem \ref{thm:sizebound} in terms of the trace norm and diamond norm.

\begin{cor}\label{cor:sizebound_trace}
Suppose the implementation $\mI=(\mH_E,\rho_E, V_{SE})$ implements channel $\Lambda_S$ with the error measured by the diamond norm: $\delta_{\mI,\diamond}^{U_S}:=\|\Lambda_{U_S}-\Lambda_S\|_\diamond$.
Let $R$ be a resource measure satisfying Properties 1, 2, 3 and $f_1$, $g_1$, $h_1$, $\mG_{U_S}$, $\mL_{U_S}$ be the functions defined in \eqref{eq:continuity}, \eqref{eq:resource generating}, \eqref{eq:resource losing} with respect to $R$ and the trace norm: $D(\rho,\sigma):=\|\rho-\sigma\|_1$. 
Then, it holds that 
\begin{align}
\mG_{U_S}+\mL_{U_S}&\le \alpha_1(\delta_{\mI,\diamond}^{U_S},d_E) + \beta_1(\delta_{\mI,\diamond}^{U_S}) 
\end{align}
where 
$\alpha_1(x,y):=f_1\left(4\sqrt{2x}\right)g_1(y) + 2f_1\left(4\sqrt{x}\right)g_1(d_S\cdot y)$ and $\beta_1(x):=h_1\left(4\sqrt{2x}\right) + 2h_1\left(4\sqrt{x}\right)$.
\end{cor}

This is a direct consequence from Theorem~\ref{thm:sizebound}, but we include a proof in the Appendixes for completeness.

\section{Relaxation of additivity condition} \label{sec:relaxation}
Although a large class of resource theories possess generic resource measures that satisfy Property 1 and Property 2, the additivity condition (Property 3) is rather a peculiar one.  
In fact, classes of resource measures that can be defined for any convex resource theory (e.g. relative entropy measure, robustness measure, convex roof measure etc.) are often only subadditive for product states. 
Thus, relaxing the additivity condition is highly desired in order for the results to be applicable to more generic scenarios.

Here, we relax the additivity condition into that for \textit{pure} product states.
It gives us much more freedom to choose resource measures because some important measures are additive only for pure product states. 
Examples for such measures include relative entropy of entanglement \cite{Vedral1997quantifying} and (logarithm of) stabilizer extent for the theory of magic~\cite{Bravyi2018simulation}, which we discuss later in detail. 

To this end, we introduce a relaxed version of Property~3 for resource measures. 
\begin{description}
 \item[Property 3']{(Additivity for pure product states) $R(\rho\otimes \sigma)=R(\rho)+R(\sigma)$ for any pure states $\rho,\ \sigma$}.
\end{description}

We also define the following resource generating/losing power for pure input states:
\begin{align}
 \mG_{U}^p &:=\max_{\ket{\psi}}\left\{R(U\dm{\psi}U^\dagger)-R(\dm{\psi})\right\}\label{eq:resource generating pure}\\
 \mL_{U}^p &:=-\min_{\ket{\psi}}\left\{R(U\dm{\psi}U^\dagger)-R(\dm{\psi})\right\}.\label{eq:resource losing pure}
\end{align}
For the same reason that $\mG_U,\ \mL_U\geq 0$, it also holds that $\mG_U^p, \mL_U^p\geq 0$ for any unitary $U$.

Then, we obtain the following trade-off relation.

\begin{thm}\label{thm:pure nogo}

Let $R$ be a resource measure satisfying Properties 1, 2, 3' and $f_L$, $g_L$, $h_L$, $\mG_{U_S}^p$, $\mL_{U_S}^p$ be the functions defined in \eqref{eq:continuity}, \eqref{eq:resource generating pure}, \eqref{eq:resource losing pure} with respect to $R$ and the Bures distance: $D(\rho,\sigma):=L(\rho,\sigma)$.    
Then, for any implementation $\mI=(\mH_E,V_{SE},\rho_E)$ with a pure state $\rho_E$, it holds that 
 \begin{align}
 &\mG_{U_S}^p+\mL_{U_S}^p\nonumber\\
 &\leq 2\left(f_L(2(1+\sqrt{2})\delta_\mI^{U_S})g_L(d_E d_S)+h_L(2(1+\sqrt{2})\delta_\mI^{U_S})\right).
 \end{align}

\end{thm}

A proof can be found in the Appendixes.
It is worth noting that $R$ does not have to be defined for general mixed states; as long as it is well-defined for pure states, the statement holds and the continuity (Property 2) can be relaxed to that for pure states. 

This Theorem leads to a variant of the aforementioned no-go theorem on perfect implementability of a resourceful unitary. 
\begin{cor}\label{cor:no-go perfect pure}
 Given the set of free unitaries $\mU_\mF$ and a finite dimensional ancillary system $\mH_E$ with $\dim\mH_E<\infty$, it is impossible to perfectly implement any unitary that can generate (or lose) nonzero resources out of pure states in terms of at least one resource measure satisfying Properties 1, 2, 3' by means of Eq. \eqref{eq:implementation with free unitary} with $\rho_E$ being a pure state.  
\end{cor}

These results encompass a standard setup where some unit resource state (e.g. Bell state for entanglement, uniform superposition state for coherence), which is usually pure, is prepared in the ancillary system. 
Although using the unit state as a resource supply appears to be more effective than using a mixed state, interestingly the requirement for Theorem~\ref{thm:pure nogo} to hold is more lenient than that for Theorem~\ref{thm:sizebound}, imposing more severe restriction on the achievable accuracy for the implementation with a pure ancillary state.

\section{Applications} \label{sec:application}

Here, we examine the validity of our results by applying them to specific physical settings. 
Although there is no systematic way of constructing a resource measure satisfying the three properties to our knowledge, it turns out that many of the important settings come with such measures tailored to each situation. 

\subsection{Systems with additive conserved quantities}
Consider a composite system consisting of subsystems $\{S_i\}_{i=1}^M$ with an observable $H_{\rm tot}=H_1\otimes \mbI^{\otimes M-1}+\mbI \otimes H_2\otimes \mbI^{\otimes M-2}+\dots$ where $H_i$ are local observables associated with subsystem $S_i$.
For these observables, we choose the set of free unitaries as the ones that conserve the expectation values for any states, or equivalently, commute with the observable. Namely, we choose 
\ba
 \mU_\mF=\lset U_{S_1\dots S_M} \sbar [H_{\rm tot}, U_{S_1\dots S_M}]=0\rset.
 \label{eq:additive conserving unitary}
\ea

An important setting that fits into this formalism is the system with conserved energy where the observable in question is the Hamiltonian of the system. Then, the free unitaries can be considered time evolutions that respect the energy conservation law, which in particular play key roles in thermodynamics on small scales \cite{Brandao2013,Horodecki2013,Aberg2013,Skrzypczyk2014,Brandao_secondlaws2015,GOUR2015informational,Tajima2016dev,Tasaki2016,Shiraishi2017,Morikuni2017,Hayashi2017}. 

For this theory, natural resource measures one can take will be the expectation value of the observable: $R(\rho_S):=\Tr[\rho_S H_S]$. 
It is clear that this measure satisfies Property 1 and 3. 
Regarding Property 2, let us take the observable of the form $H_S = \sum_{j=0}^{d_S-1} j\dm{j}$. Then, we get  
\bal
 |R(\rho)-R(\sigma)|&= |\Tr[(\rho-\sigma)H_S]|\\
 &= |\sum_j (\rho_{jj}-\sigma_{jj})H_{S,j}|\\
 &\leq \sum_j |(\rho_{jj}-\sigma_{jj})| |H_{S,j}|\\
 &\leq \sum_j |(\rho_{jj}-\sigma_{jj})| \|H\|_\infty \\
 &= \|\Delta(\rho-\sigma)\|_1 (d_S-1)\\
 &\leq \|\rho-\sigma\|_1 (d_S-1)
 \label{eq:continuity_energy}
\eal
where $\rho_{jj}=\bra{j}\rho\ket{j}$, $\sigma_{jj}=\bra{j}\sigma\ket{j}$, $H_{S,j} = \bra{j}H_S\ket{j}$, $\Delta$ is the dephasing with respect to the eigenbasis of $H_S$, and we used the contractivity of the trace norm under CPTP maps in the last inequality. 
Thus, for this case one can take $f_1(x)=x$, $g_1(x)=x$, and $h_1(x)=-x$ in Corollary~\ref{cor:sizebound_trace}, and we conclude that the finite dimensional environment does not allow for perfect implementation of unitary that changes the energy by any energy-conserving unitary and an energy ``battery" state, which qualitatively reproduces the results in \cite{Tajima2018uncertainty,Tajima2019coherence}.
Although we considered the observable with uniform spectrum, a similar argument can be applied to other observables with more general form.

It will be worth pointing out that this is a situation where our approach in which one does not necessarily need to assume any underlying resource theory becomes useful, since the concept of free states and free operations for this setting can be ambiguous --- from the perspective that the energy is resource, one could say that the ground state $\ket{0}$ is free, but in that case the set of free unitaries defined in terms of free operations does not coincide with the set of energy-conserving unitaries since any unitary that can change energy but does not affect the ground state (e.g. bit flip between $\ket{1}$ and $\ket{2}$) also becomes free in this definition. 
Thus, when the focus is put on the conservation law, it is natural to just consider the set of free unitaries that meets the physical requirement. 

On the other hand, by shifting our focus on the type of resource of interest from the expectation value of the observable to that of \textit{fluctuation}, the underlying resource theory can be naturally identified as the resource theory of asymmetry \cite{Gour2008,Marvian2016}. 
In particular, the resource theory of asymmetry with U(1) group with unitary representation $U_t=e^{iH_S t}$ is equipped with a family of resource monotones that are additive for product states known as metric-adjusted skew informations \cite{Hansen2008,Zhang2017,Takagi2018skew}. 
One of the examples in this family is the well-known Wigner-Yanase skew information \cite{Wigner1963,marvian2014extending} defined as 
\bal
  I^{WY}(\rho,H_S) &= -\frac{1}{2}\Tr([\sqrt{\rho},H_S]^2)\\
  &= \Tr(\rho H_S^2) - \Tr(\sqrt{\rho}H_S\sqrt{\rho}H_S).
\eal
Since this satisfies Property 1 and 3, Theorem \ref{thm:sizebound} and Corollary \ref{cor:no-go perfect} can be applied with respect to this measure as well, providing another way of looking at the trade-off relation. 

Finally, when the observable of interest is the Hamiltonian, the free unitaries in \eqref{eq:additive conserving unitary} preserve the Gibbs state $\tau=\exp(-H_S/T)/Z$ where $T$ is the temperature and $Z$ is the partition function of the system. 
This motivates us to consider the ``athermality", a measure indicating the distance from the Gibbs state to the given state, and especially the free energy is recovered by taking the relative entropy as a distance measure: 
\ba
 A_R(\rho):=S(\rho||\tau)=\frac{1}{T}(F(\rho)-F(\tau))
\ea
where $F(\rho):=\Tr[\rho H_S]-T S(\rho)$ is the free energy. 
It is then easy to see that this also satisfies all the three properties. 

\subsection{Coherence}

Consider the theory of coherence \cite{aberg2006quantifying,Baumgratz2014,Streltsov2017} where one is interested in the degree of superposition with respect to the given preferred basis $\{\ket{i}\}$.
For this theory, the set of incoherent states $\mathscr{I}:={\rm conv}(\{\dm{i}\})$ is a reasonable choice for the free states, and one can naturally choose the relevant free unitaries $\mU_\mF(d)=\lset U \sbar U=\sum_{j=0}^{d-1} e^{i\theta_j}\ketbra{\pi(j)}{j}\rset$ where $\pi$ is the permutation on $\{0,\dots,d-1\}$, which is often called the set of incoherent unitaries.  

As a resource measure, let us consider a standard coherence measure, the relative entropy of coherence: 
\ba
 C_R(\rho):= \min_{\sigma\in\mathscr{I}}S(\rho||\sigma) = S(\Delta(\rho))-S(\rho).
 \label{eq:rel ent coherence}
\ea

For this measure, it is easy to see that Property 1 is satisfied.
The explicit form of $C_R$ in \eqref{eq:rel ent coherence} ensures Property~3 as well because of the additivity of the von Neumann entropy for product states. 
As for Property 2, recall the following asymptotic continuity property that holds for relative entropy measure $M_R(\rho):=\inf_{\sigma\in\mF}S(\rho||\sigma)$ with $\mF$ being any convex and closed set of positive semidefinite operators that contains at least one full-rank operator~\cite{Winter2016asymptotic}:
\bal
 |M_R(\rho)-M_R(\sigma)|\leq \kappa \epsilon + (1+\epsilon)b\left(\frac{\epsilon}{1+\epsilon}\right)
 \label{eq:asymp. continuity}
\eal
for any two states $\frac12\|\rho-\sigma\|_1\leq \epsilon$ where $\kappa:=\sup_{\tau,\tau'}\{M_R(\tau)-M_R(\tau')\}$ and $b(x):=-x\log x -(1-x)\log (1-x)$ is the binary entropy.
For the case of theory of coherence, \eqref{eq:asymp. continuity} reduces to the following bound:
\bal
 |C_R(\rho)-C_R(\sigma)|\leq \epsilon\log d  + (1+\epsilon)b\left(\frac{\epsilon}{1+\epsilon}\right),
\eal
for which we find $f_1(x)=x$, $g_1(x)=\log x$, and $h_1(x)=(1+x)b(x/(1+x))$.
Since this measure is also faithful, i.e. $C_R(\rho)=0$ iff $\rho\in\mathscr{I}$, Corollary \ref{cor:sizebound_trace} implies that any coherence generating unitary that can create a coherent state out of an incoherent state cannot be implemented with zero-error with the aid of any coherent state acting on a finite-dimensional ancillary system. 

\subsection{Entanglement}
Arguably, entanglement is one of the most important resources to consider, which has a strong connection to operational tasks in quantum information processing. 
In particular, using only local operations and classical communication to implement desired global operations with the help of preshared entanglement is a key idea of quantum network and distributed quantum computing \cite{Duan2010qnetwork,Pirker2018qnetwork}, and methodology as well as necessary entanglement cost for implementing global gates with local operations and classical communication have been considered for various settings \cite{Eisert2000optimal,Soeda2011entanglement,Chen2014nonlocal,Chen2016entanglement,Wakakuwa2017coding,Wakakuwa2019complexity,Pirandola2017fundamental}. 
Our formalism addresses a more restricted scenario where the parties only have access to local gates in order to implement a desired global gate with the aid of preshared entanglement.   
Our results induce necessary size of the shared entangled state and imply the impossibility of perfectly implementing any entangling gate with finite-sized aiding system. 
Since it is clearly possible to perfectly implement any global unitary if classical communication is allowed (via quantum teleportation), our results clarify the significance of classical communication for the situations such as distributed quantum computing~\footnote{One could alternatively argue the impossibility of implementing entangling gates only using local operations and shared entanglement by noting that no signaling is allowed under this setup.}. 

In order to apply our results, we need to find an entanglement measure satisfying the three properties. In particular, one needs to be careful about the additivity property since some well-known entanglement measures (e.g. such as the (max-)relative entropy of entanglement \cite{Vollbrecht2001rel,Datta2009max}, robustness of entanglement \cite{Vidal1999robustness}) are only subadditive even for product states, and it had been indeed an important program to find an additive measure of entanglement. 
As a result, the squashed entanglement was introduced as an additive entanglement measure \cite{Christandl2004squashed}, and its continuity was also shown \cite{Alicki_2004fannes}.  
In addition, the conditional entanglement of mutual information~\cite{Yang2008conditional} was introduced as another additive and continuous measure of entanglement.
Remarkably, this measure can be easily extended to multipartite entanglement, which allows our results to be applied to the multipartite scenarios. 

On the other hand, Theorem~\ref{thm:pure nogo} allows us to avoid this subtlety and take an even simpler entanglement measure. 
For instance, the relative entropy of entanglement is additive for pure product states, as can be seen by noting that it reduces to the entanglement entropy for pure states. Since it clearly satisfies Property 1 and 2 as well, Theorem \ref{thm:pure nogo} and Corollary \ref{cor:no-go perfect pure} immediately follows for such measure.

\subsection{Fault-tolerant quantum computation}
To realize the quantum computation in a noise-resilient fashion, which is so called \textit{fault-tolerant quantum computation} \cite{Shor1996fault,Preskill1998fault}, encoding quantum states into quantum error correcting codes and carrying out logical computation inside the code space is essential. 
Since many promising error correcting codes allow for relatively efficient implementation of the logical Clifford gates in a fault-tolerant manner \cite{Steane1996error,Shor1996scheme,Steane1997active,Fowler2012surface,Bombin2006topological}, for the situations where those codes are in use, Clifford gates can be naturally considered ``free". 
However, since Clifford gates do not form a universal gate set, some non-Clifford gate needs to be implemented fault-tolerantly, and a popular way of realizing it is via the gate teleportation \cite{Gottesman1999demonstrating}, in which ``magic states" \cite{Bravyi2005universal} are injected as resources of ``non-Cliffordness". 
Since good logical magic states are hard to prepare in general, a magic-state distillation protocol \cite{Bravyi2005universal} should be run beforehand to increase the quality of the noisy magic states. 
However, a large overhead cost comes with the distillation protocols and how to reduce the overhead has been under active research \cite{Bravyi2012magic,Fowler2013surface,Jones2013multilevel,Duclos-Cianci2013distillation,Duclos-Cianci2015reducing,Campbell2017unified,O'Gorman2017quantum,Haah2018codesprotocols,Campbell2018magicstateparity,Fowler2018low,Gidney2019efficientmagicstate,Litinske2019magic} (error correcting codes that avoid using the magic-state distillation have also been investigated \cite{Paetznick2013universal,Anderson2014fault,Bombin2015gauge,JOchym2014using,Nikahd2017nonuniform,Chamberland2016threshold,Yoder2016universal,Takagi2017error}), and this costly nature of magic states motivates us to consider the resource theory of magic, which considers the ``magicness" as precious resources. 

The resource theory of magic is defined by the set of free states called stabilizer states, which is the convex combinations of pure states produced by Clifford gates~\cite{Veitch_2014stab}.
By definition, non-Clifford gates are able to create non-stabilizer sates out of stabilizer states, and as described above it is an essential building block for universal quantum computation. 
This operationally motivated framework leads us to a natural question on how well a non-Clifford gate could be implemented by Clifford gates with the aid of magic states as resources. 
Our results address this question by considering appropriate resource measures for magicness. 
We consider the cases of qubits (dimension 2) and quopits (qudits with odd-prime dimensions) separately. 

\subsubsection{Qubits}
Although one can consider valid magic monotones defined for multiqubit states (e.g. relative entropy of magic \cite{Veitch_2014stab}, robustness of magic \cite{howard_2017}), they are not additive for product states in general, which prevents us from applying Theorem \ref{thm:sizebound}.
However, Theorem \ref{thm:pure nogo} turns out to be useful in this case since there indeed exists a measure defined for pure states and additive for pure product multiqubit states. 
To this end, consider the \textit{stabilizer extent} introduced in \cite{Bravyi2018simulation}:
\bal
 \xi(\ket{\psi}):= \min\lset \left(\sum_i |c_i|\right)^2 \sbar \ket{\psi}=\sum_i c_i \ket{\phi_i}\rset 
\label{eq:stabilizer extent}
\eal
where $\ket{\phi_i}$ are pure stabilizer states.
The stabilizer extent was originally introduced for investigating the overhead cost for classically simulating quantum circuits, but we find that it is also useful for our purpose, providing another perspective to this measure.
Let us take our resource measure as $R(\dm{\psi})=\log \xi(\ket{\psi})$.
It was shown that the stabilizer extent is multiplicative for tensor products between states supported on up to three qubits \cite{Bravyi2018simulation}, and thus $R$ satisfies Property 3'.  
Property 1 is also satisfied because of the monotonicity of $\xi$ under Clifford gates and reversibility of Clifford unitary under another Clifford unitary  (since Clifford gates constitute a group).
As for Property 2, we first remark that our measure coincides with the \textit{max-relative entropy of magic} for pure states as shown in Ref.~\cite{regula_2018}, where the max-relative entropy measure is defined as
\bal
 \mfD_{\rm max}(\rho) := \min\lset r \sbar\rho \cleq 2^r \sigma,\ \sigma\in{\rm STAB}\rset 
\label{eq:dmax magic}
\eal
where STAB refers to the set of stabilizer states, and $\cleq$ denotes the inequality with respect to the positive semidefiniteness.  
Then, we prove the following continuity bound for max-relative entropy of magic, which may be of independent interest. 
Using the identity between $R$ and \eqref{eq:dmax magic} for pure states, the continuity of stabilizer extent is derived as a special case of this result. 
It would be also worth noting that the following result holds for the max-relative entropy measure defined for any convex resource theory that includes the maximally mixed state as a free state. (One can also easily extend the relation to the theories with at least one full-rank free state.) 

\begin{pro}\label{pro:continuity dmax magic}
 Let $\rho,\sigma\in \mD(\mH_{d_S})$ and suppose that $\|\rho-\sigma\|_1< 1/(2d_S)$. Then, it holds that
 \bal
  |\mfD_{\max}(\rho)-\mfD_{\max}(\sigma)|\leq 2\|\rho-\sigma\|_1 d_S.
 \eal
\end{pro}

The proof is presented in the Appendixes.
Our results provide an interesting implication for implementation of non-Clifford gates. 
Suppose we are given qubits acting on system $A$ and try to implement some non-Clifford gate $U_{\rm NC}$ on the subsystem $A_1\subset A$ by applying Clifford gates on $A$. 
Let $N$ be the number of qubits supported on the subsystem $A\setminus A_1$. 
Then, our results imply that in order to realize the implementation accuracy $\epsilon$ with respect to the diamond norm, the required number of qubits $N$ must scale as $\Omega\left(\log\left(\frac{\mG_{U_{\rm NC}}^p+\mL_{U_{\rm NC}}^p}{\sqrt{\epsilon}}\right)\right)$.
This observation explicitly tells us the importance of measurement + feedforward (adaptive) operations for quantum circuits to gain their power.

\subsubsection{Quopits}
For the case when the dimension of the system that each qudit acts on is odd-prime, ``mana" was introduced as a magic monotone~\cite{Veitch_2014stab}:
\ba
 \mM(\rho):=\log \left(\sum_{\bf u}|W_{\rho}({\bf  u})|\right)
\ea
where $W_{\rho}({\bf u})$ is the discrete Wigner function for state $\rho$ \cite{Gross2006hudson}. 
The mana essentially measures the total negativity of the discrete Wigner function, which is motivated by the fact that stabilizer states only take non-negative value for the discrete Wigner function.
An important property of this measure for our purpose is that it is additive for product states, which comes from that the discrete Wigner function for a product state is just the multiplication of the two discrete Wigner functions of the states that constitute the product state. 
It is also continuous (although it is not asymptotically continuous as shown in \cite{Veitch_2014stab}), and Property~1 can be also easily seen by the monotonicity of mana under Clifford gates and the fact that the application of Clifford gate can be reversed by another Clifford gate. 
Thus, Theorem \ref{thm:sizebound} and Corollary \ref{cor:no-go perfect} can be applied with respect to the mana measure. 

Note that the mana is \textit{not} faithful: there exists a magic state $\rho$ with $\mM(\rho)=0$~\cite{Veitch_2012bound}. However, the discrete Hudson's theorem~\cite{Gross2006hudson} ensures that it is faithful for pure states, which is enough to show that any non-Clifford unitary cannot be implemented with zero-error with a finite number of magic states.

\section{Toward full generality} \label{sec:full generality}

Although Theorem \ref{thm:pure nogo} covers most of the known important settings, one could still argue that some theory of interest may not come with a resource measure that satisfies all the three properties, especially the additivity condition. 
Here, we focus on the qualitative no-go statement and see that it is quite unlikely for the perfect implementation of resourceful unitary to be possible even in more general settings. 

Suppose that free unitary $V_{SE}$ and pure state $\ket{\phi}$ allow for an exact implementation of $U_S$, i.e. $\Tr_E\left[V_{SE}(\rho_S\otimes \dm{\phi}_E)V_{SE}^\dagger\right]=U_S\rho_S U_S^\dagger$ for any $\rho_S$. 
By taking $\delta_\mI^{U_S}=0$ in \eqref{eq:no correlationpure}, we get
\ba
 \Tr_S\left[V_{SE}(\rho_S\otimes \dm{\phi}_E)V_{SE}^\dagger\right]=\sigma_E'
\ea
where $\sigma_E'$ is a pure state. 
Since states with pure reduced states are only product states, we know that the total state must look like 
\ba
 V_{SE}(\rho_S\otimes \dm{\phi}_E)V_{SE}^\dagger = U_S\rho_S U_S^\dagger\otimes \sigma_E'.
\label{eq:perfect implementation product form}
\ea
Then, we get for any $\rho_S$ and any measure $R$ that is invariant under free unitaries that
\bal
 R(\rho_S\otimes \dm{\phi}) &= R(V_{SE}(\rho_S\otimes \dm{\phi})V_{SE}^\dagger) \\
 &= R(U_S\rho_S U_S^\dagger\otimes \sigma_E')
 \label{eq:no-go cond}
\eal
Thus, for the given theory, unless \textit{any} resource measure with Property 1 (but not necessarily Property 2, 3, 3') satisfies \eqref{eq:no-go cond} for \textit{any} $\rho_S$, it is impossible to implement the target $U_S$ exactly.
Note that this is a very strong restriction, and when $R$ is additive for product states, Corollaries \ref{cor:no-go perfect} and \ref{cor:no-go perfect pure} are reproduced. 

Let us impose another natural condition on $R$ that it be a subadditive monotone for some resource theory in which composition of free states and partial trace are free operations. 
For such cases, one can show that $R(\dm{\phi})=R(\sigma_E')$ as follows. 
Take a free state $\tau_S$ and $\eta_S=U_S^\dagger \tau_S U_S$. Then, we get
\bal
 R(\dm{\phi}) &\geq R(\tau_S\otimes \dm{\phi}) \\
 &= R(U_S\tau_S U_S^\dagger \otimes \sigma_E')\geq R(\sigma_E')
\eal
and
\bal
 R(\sigma_E') &\geq R(U_S\eta_S U_S^\dagger\otimes \sigma_E') \\
 &= R(\eta_S \otimes \dm{\phi})\geq R(\dm{\phi}).
\eal
where to show both of the above relations we used that the composition of free states is a free operation in the first inequalities, the invariance of $R$ under free unitaries and \eqref{eq:perfect implementation product form} in the equalities, and that the partial trace is a free operation in the last inequalities together with the assumption that $R$ is a monotone under free operations.   

This makes it even more surprising that Eq. \eqref{eq:no-go cond} holds for any $\rho_S$ for resourceful unitary $U_S$ since it would indicate that attaching ancillary states with the same amount of resources to two states with different amount of resources would necessarily produce the states with the same amount of resources. 
We leave the thorough analysis on how general the no-go statement can be made for future work. 

\section{Conclusions} \label{sec:conclusions}

We considered a general setting where one aims to implement a target unitary with access to a restricted set of unitaries as well as ancillary system.
We derived a trade-off relation between the implementation accuracy and the size of the ancillary system in terms of the amount of the resources that can be changed by the target unitary with respect to resource measures that satisfy three properties: invariance under free unitaries, continuity, and additivity for product states. 
Using this relation, we presented a fundamental no-go theorem on the perfect implementation of resourceful unitaries with finite-dimensional ancillary systems. 
We further relaxed the subtle condition in the above three properties, additivity for product states, and showed an analogous trade-off relation that only requires the resource measures to be additive for pure product states, in addition to the other two properties.  
We exemplified the wide validity of our results by applying them to various important settings and discussed the physical significance implied by the results for specific settings. We finally discussed the feasibility of extending our no-go results to even more general settings that do not assume all the properties for the resource measures we considered. 

For future work, it will be intriguing to clarify whether some of the required properties for resource measures considered in this work can be dropped to obtain a similar trade-off relation.  
It will also be interesting to investigate how good our lower bounds are in general by constructing upper bounds with explicit protocols that approximately implement desired unitaries. 
 
\textit{Note added}. --- Recently, we became aware of the independent related work by Chiribella, Yang, and Renner \cite{Chiribella2019energy}. 

\begin{acknowledgments}
We thank Tomoyuki Morimae for fruitful discussions. R. T. acknowledges the support of NSF, ARO, IARPA, and the Takenaka Scholarship Foundation. H. T. acknowledges the support of JSPS (Grants-in-Aid for Scientific Research No. JP19K14610). 
\end{acknowledgments}


\appendix

\section{Proof of Theorem \ref{thm:sizebound}}

We first retrieve the main lemma we use for the readers' convenience.

\begin{lem}[No-correlation lemma \cite{Tajima2019coherence}]\label{lem:no-correlation}
Let $\Lambda_{AB}$ be a channel on the composite system $AB$ and $U_A$ be a unitary operation on $A$.
We consider three possible initial states of $A$: $\rho^{(0)}_{A}$, $\rho^{(1)}_{A}$, and $\rho^{(0+1)}_{A}:=(\rho^{(0)}_{A}+\rho^{(1)}_{A})/2$ and write the initial state of $B$ as $\rho_{B}$.
We refer to the final states of $AB$ and $B$ with the initial state $\rho^{(i)}_{A}$ ($i=0,1,0+1$) as 
\begin{align}
\sigma^{(i)}_{AB}&:=\Lambda_{AB}(\rho^{(i)}_{A}\otimes\rho_{B}),\\
\sigma^{(i)}_{B}&:=\Tr_{A}[\sigma^{(i)}_{AB}].
\end{align}
Let $\Lambda_A$ be the channel implemented by the implementation $\mI=(\mH_E,\Lambda_{AB},\rho_B)$, i.e. $\Lambda_{A}(\cdot):=\Tr_{B}[\Lambda_{AB}(\cdot\otimes\rho_B)]$ and write the accuracy of implementation of $U_A$ with implementation $\mI$ for input state $\rho_A^{(i)}$ as $\delta_\mI^{U,(i)}:=\delta_\mI^{U}(\rho_A^{(i)})$ as in \eqref{eq:accuracy bures}.
Then, for any $U_A$ and $\mI$, we have the following relations: $\\$
1. It holds that 
\begin{align}
L(\sigma^{(i)}_{AB}, U_{A}\rho^{(i)}_{A}U^{\dagger}_{A}\otimes\sigma^{(i)}_{B})\le2\delta_\mI^{U_A,(i)}.\label{eq:no correlation}
\end{align}$\\$
2. There exists a state $\sigma'^{(0+1)}_{B}$ of $B$ such that
\begin{align}
L(\sigma^{(0)}_{B},\sigma'^{(0+1)}_{B})+L(\sigma'^{(0+1)}_{B},\sigma^{(1)}_{B})\le2\sqrt{2}\delta_\mI^{U_A,(0+1)}.\label{eq:no correlation int state}
\end{align}
Moreover, if $\rho_B$ is a pure state and $\Lambda_{AB}$ is a unitary operation, one can take a pure state for $\sigma'^{(0+1)}_{B}$.
\end{lem}

We are now in a position to prove Theorem \ref{thm:sizebound}.
\begin{proof}
Define $\rho_S^{(i)},\ i=0,1$ as 
\begin{align}
\rho^{(0)}_S&:=\mbox{argmax} (R(U_S\rho_SU^{\dagger}_S)-R(\rho_S))\\
\rho^{(1)}_S&:=\mbox{argmin} (R(U_S\rho_SU^{\dagger}_S)-R(\rho_S))
\end{align}
and corresponding final states on $SE$ and $E$ as 
\begin{align}
\sigma^{(i)}_{SE}&:=V_{SE}(\rho^{(i)}_S\otimes\rho_E)V_{SE}^{\dagger},\\
\sigma^{(i)}_{E}&:=\Tr_S[\sigma^{(i)}_{SE}].
\end{align}
Due to Property 1 and 3 of the resource measure $R$, we have
\begin{align}
R(\rho^{(i)}_S)+R(\rho_E)=R(\sigma^{(i)}_{SE})\label{14}.
\end{align}
Using \eqref{eq:no correlation}, we get
\begin{align}
L(\sigma^{(i)}_{SE},U_S\rho^{(i)}_SU^{\dagger}_S\otimes\sigma^{(i)}_{E})\le2\delta_\mI^{U_S}.\label{15}
\end{align}
Due to Property 2 of $R$ and \eqref{14}, \eqref{15}, we obtain
\begin{align}
\begin{aligned}
&|R(\rho^{(i)}_{S})+R(\rho_E)-R(U_S\rho_S^{(i)}U^{\dagger}_S)-R(\sigma^{(i)}_E)|\\
&\hspace{1cm}\le f_L(2\delta_\mI^{U_S})g_L(d_Ed_S)+h_L(2\delta_\mI^{U_S}).
\end{aligned}
\label{eq:continuity initial final each}
\end{align}
Using the triangle inequality and \eqref{eq:continuity initial final each}, we get 
\begin{align}
\begin{aligned}
&|R(\rho^{(0)}_{S})-R(U_S\rho_S^{(0)}U^{\dagger}_S)-R(\sigma^{(0)}_E)\\
&-R(\rho^{(1)}_{S})+R(U_S\rho_S^{(1)}U^{\dagger}_S)+R(\sigma^{(1)}_E)|\\
&\hspace{1cm}\le 2\left(f_L(2\delta_\mI^{U_S})g_L(d_Ed_S)+h_L(2\delta_\mI^{U_S})\right).
\end{aligned}
\end{align}
Another use of the triangle inequality leads to
\begin{align}
&|R(\sigma^{(0)}_E)-R(\sigma^{(1)}_{E})|\nonumber\\
&\ge |R(U_S\rho^{(0)}_SU^{\dagger}_S)-R(\rho^{(0)}_S)-R(U_S\rho^{(1)}_SU^{\dagger}_S)+R(\rho^{(1)}_S)|\nonumber\\
&\ -2\left(f_L(2\delta_\mI^{U_S})g_L(d_Ed_S)+h_L(2\delta_\mI^{U_S})\right)\nonumber\\
&=\mG_{U_S}+\mL_{U_S}-2\left(f_L(2\delta_\mI^{U_S})g_L(d_Ed_S)+h_L(2\delta_\mI^{U_S})\right)
\label{eq:remaining resource bound 1}
\end{align}
where we used $\mG_{U_S},\mL_{U_S}\geq 0$ in the equality.
On the other hand, using \eqref{eq:no correlation int state} together with triangle inequality and Property 2 of $R$, we get
\begin{align}
&|R(\sigma^{(0)}_E)-R(\sigma^{(1)}_{E})|\nonumber\\
&\ \le f_L(2\sqrt{2}\delta_\mI^{U_S})g_L(d_E)+h_L(2\sqrt{2}\delta_\mI^{U_S}).
\label{eq:remaining resource bound 2}
\end{align}
Combining \eqref{eq:remaining resource bound 1} and \eqref{eq:remaining resource bound 2}, we finally obtain
\begin{align}
\mG_{U_S}+\mL_{U_S}&\le f_L(2\sqrt{2}\delta_\mI^{U_S})g_L(d_E)+h_L(2\sqrt{2}\delta_\mI^{U_S})\nonumber\\
&\ +2\left(f_L(2\delta_\mI^{U_S})g_L(d_Ed_S)+h_L(2\delta_\mI^{U_S})\right)\nonumber\\
&=\alpha_L(\delta_\mI^{U_S},d_E)+\beta_L(\delta_\mI^{U_S}).
\end{align}
\end{proof}

\section{Proof of Corollary \ref{cor:sizebound_trace}}
\begin{proof}
Recall the relation between the Bures distance and the trace distance \cite{Fuchs1999distance} 
\ba
 \frac{1}{2}\left(L(\rho,\sigma)\right)^2\leq \frac{1}{2}\|\rho-\sigma\|_1\leq L(\rho,\sigma),
\ea
which also implies $\delta_{\mI}^{U_S}\leq \sqrt{\delta_{\mI,\diamond}^{U_S}}$. 
Then, \eqref{eq:no correlation} and \eqref{eq:no correlation int state} imply 
\ba
 \frac{1}{2}\|\sigma_{SE}^{(i)} - U_S\rho_S^{(i)}U_S^\dagger\otimes\sigma_E^{(i)}\|_1\leq 2\sqrt{\delta_{\mI,\diamond}^{U_S}}
\ea
and
\ba
 \frac{1}{2}\|\sigma_{B}^{(0)} - \sigma_{B}^{(1)}\|_1\leq 2\sqrt{2\delta_{\mI,\diamond}^{U_S}}
\ea
Then, the same proof as Theorem \ref{thm:sizebound} can be employed to obtain the statement. 
\end{proof}

\section{Proof of Theorem~\ref{thm:pure nogo}}
\begin{proof}
Lemma \ref{lem:no-correlation} together with the assumption that $\rho_E$ is pure ensures that there exists a pure state $\sigma_E'$ that satisfies \eqref{eq:no correlation int state}, namely
\bal
 L(\sigma_E^{(i)},\sigma_E')&\leq L(\sigma_E^{(0)},\sigma_E') + L(\sigma_E^{(1)},\sigma_E') \\
 &\leq 2\sqrt{2}\delta_\mI^{U_S} \label{eq:no correlationpure}.
\eal

Then, we obtain
\begin{align}
&L(\sigma^{(i)}_{SE},U_S\rho^{(i)}_SU^{\dagger}_S\otimes\sigma_E')\nonumber\\
&\leq L(\sigma^{(i)}_{SE},U_S\rho^{(i)}_SU^{\dagger}_S\otimes\sigma_E^{(i)})\nonumber\\
&\ \ +L(U_S\rho^{(i)}_SU^{\dagger}_S\otimes\sigma_E^{(i)},U_S\rho^{(i)}_SU^{\dagger}_S\otimes\sigma_E')\nonumber\\
&\leq 2\delta_\mI^{U_S} + L(\sigma_E^{(i)},\sigma_E')\nonumber\\
&\leq 2(1+\sqrt{2})\delta_\mI^{U_S} \label{eq:no-correlation pure thm proof}
\end{align}
where in the first inequality we used the triangle inequality, in the second inequality we used \eqref{eq:no correlation} and the fact that $L(\rho\otimes \sigma, \rho\otimes \tau)=L(\sigma,\tau)$, and in the third inequality we used \eqref{eq:no correlationpure}. 

Let $\rho_S^{(0)}$ and $\rho_S^{(1)}$ be pure states that achieve \eqref{eq:resource generating pure} and \eqref{eq:resource losing pure} respectively.  
Then, Property 1 and 3' of $R$ lead to 
\bal
 R(\sigma_{SE}^{(i)})=R(\rho_S^{(i)})+R(\rho_E).
 \label{eq:invariance unitary pure}
\eal
and 
\bal
 R(U_S\rho^{(i)}_SU^{\dagger}_S\otimes\sigma_E')=R(U_S\rho_S^{(i)}U_S^\dagger)+R(\sigma_E').
 \label{eq:additive pure}
\eal
Combining Property 2, \eqref{eq:no-correlation pure thm proof}, \eqref{eq:invariance unitary pure}, \eqref{eq:additive pure}, we get
\begin{align}
&|R(\rho^{(i)}_{S})+R(\rho_E)-R(U_S\rho_S^{(i)}U^{\dagger}_S)-R(\sigma_E')|\nonumber\\
&\le f_L(2(1+\sqrt{2})\delta_\mI^{U_S})g_L(d_E d_S)+h_L(2(1+\sqrt{2})\delta_\mI^{U_S}).
\end{align}
Hence, 
\begin{align}
0 &=R(\rho_E)-R(\sigma_E') + R(\sigma_E')-R(\rho_E)\nonumber\\
&\ge R(U_S\rho^{(0)}_SU^{\dagger}_S)-R(\rho^{(0)}_S)-R(U_S\rho^{(1)}_SU^{\dagger}_S)+R(\rho^{(1)}_S)\nonumber\\
&\ -2\left(f_L(2(1+\sqrt{2})\delta_\mI^{U_S})g_L(d_E d_S)+h_L(2(1+\sqrt{2})\delta_\mI^{U_S})\right)\nonumber\\
&=\mG_{U_S}+\mL_{U_S}\nonumber\\
&\ -2\left(f_L(2(1+\sqrt{2})\delta_\mI^{U_S})g_L(d_E d_S)+h_L(2(1+\sqrt{2})\delta_\mI^{U_S})\right),
\end{align}
which proves the statement.
\end{proof}

\section{Proof of Proposition \ref{pro:continuity dmax magic}}
\begin{proof}
We assume $\mfD_{\max}(\rho)\geq\mfD_{\max}(\sigma)$ without loss of generality.
The definition of max-relative entropy measure \eqref{eq:dmax magic} admits the following dual form \cite{boyd_2004}:
\begin{equation}\begin{aligned}
{\text{\rm maximize }}& \ \  \log \Tr[\rho X]  \\
{\text{\rm subject to }}&\ \ X\cgeq 0\\
&\ \ \Tr[\tau X] \leq 1,\ \forall \tau\in {\rm STAB}.
\end{aligned}
\label{eq:dmax magic dual}
\end{equation}

Let $X_{\rho}$ be an optimal solution that achieves \eqref{eq:dmax magic dual} for state $\rho$. 
Then, we obtain
\bal
 \mfD_{\max}(\sigma)&\geq \log \Tr[\sigma X_\rho]\\
 &\geq \log \left(\Tr[\rho X_\rho]-\|\rho-\sigma\|_1\|X_\rho\|_\infty\right)\\
 &= \mfD_{\max}(\rho)+\log\left(1-\frac{\|\rho-\sigma\|_1\|X_\rho\|_\infty}{\Tr[\rho X_\rho]}\right)\\
 &\geq \mfD_{\max}(\rho)+\log\left(1-\|\rho-\sigma\|_1d_S\right)\\
 & \geq\mfD_{\max}(\rho)-2\|\rho-\sigma\|_1d_S
\label{eq:continuity proof}
\eal
The first inequality is because $X_\rho$ is a suboptimal solution for $\sigma$. 
The second inequality is because of the same argument in \eqref{eq:continuity_energy}.
The third inequality is because it holds that $\|X_\rho\|_\infty\leq d_S$ from the second constraint in \eqref{eq:dmax magic dual} together with the fact that the maximally mixed state $\mbI/d_S$ is a stabilizer state, and that $\Tr[\rho X_\rho]\geq 1$ because $\mbI$ serves as a suboptimal solution for $X$ that gives $\Tr[\rho \mbI]=1$.
The fourth inequality is because it holds that $\log(1-x)\geq -2x$ for $0\leq x\leq 1/2$ (note that we take the base 2 for the logarithm), where we used the assumption that $\|\rho-\sigma\|_1< 1/(2d_S)$. 
Note also that the logarithm in \eqref{eq:continuity proof} is always well-defined because $\Tr[\rho X_\rho]\geq 1$ and $\|\rho-\sigma\|_1\|X\|_\infty\leq 1/2$.
The statement is reached by combining the assumption that $\mfD_{\max}(\rho)\geq \mfD_{\max}(\sigma)$.
\end{proof}

\bibliographystyle{apsrmp4-2}
\bibliography{myref}

\end{document}